\documentclass{llncs}

\usepackage[hyphens]{url}

\usepackage{ifpdf}
 \ifpdf
\usepackage[pdftex]{graphicx}
 \else
\usepackage[dvips]{graphicx}
 \fi

\fussy

\begin{document}

\title{
Crypto Currency Regulation
and Law Enforcement Perspectives
}

\author{
Nicolas T. Courtois\inst{1} \and Kacper T. Gradoń\inst{2}\inst{,}\inst{3} \and Klaus Schmeh\inst{4}
}

\institute{
$^{1}$ Department of Computer Science, University College London, UK\\
$^{2}$ Faculty of Law and Administration, University of Warsaw, Poland\\
$^{3}$ Department of Security and Crime Science, University College London, UK\\
$^{4}$ Cipherbrain.net, Gelsenkirchen, Germany
}

\maketitle

\abstract{
This paper provides an overview of how crypto currency and blockchain engineering interacts with the law enforcement.
We point out that a large proportion of crypto users are amateur investors and
the dominant and the largest segment in crypto crime are simply investment scams (!).
We look at various questions of criminal use and misuse of technology,
especially in the areas of money laundering or cashing out the profits originating from illicit activities.
The aim of the paper is to raise a set of concerns arising in the criminal justice and policing circles,
based on the interviews with law enforcement practitioners,
and to see how cryptos could be reconciled with public security and safety.
We propose a simplified classification of crimes related to crypto currency.
We study the development of blockchains in a broader context of applied cryptography and payment technology. 
Ransomware is a big threat but we also need protection against corporate misconduct or negligence,
with untested financial services breaching customer trust or government regulations.
Not paying taxes is illegal, but there is more at stake:
exposing crypto holders to losing all their savings in scams or thefts.
Interestingly, privacy helps to defend on multiple fronts:
against social engineering, targeted crime, scams,
and also against cybersecurity thefts and hacks. 
}


\section{
{Introduction}}
\label{sec:1:intro}



Blockchain and associated payment methods are the emerging technologies,
which have 220 million of users \cite{CryptoComAdoptReport0721}.
In our view, the current crypto landscape and ecosystem must be considered to be more than
just an evolution of an electronic payment system invented by a certain Satoshi Nakamoto \cite{cg02}.
It has then in fact “diversified” from the initial intended function of transferring coins between peers. 
Crypto and blockchain technology is now becoming a major disruptor of how our economy and our society functions.
The challenge goes far beyond small niche sectors such as trans-border foreign currency activity.
It is now entering substantially larger parts of the economy:
like competing with banks and potentially also stock markets,
on attracting investor deposits and investment activity fees.
It is a powerful programmable toolset capable of doing so many things in finance and beyond \cite{cg48}.
Many of these are new and  entirely impossible to do without cryptography
and some form of decentralized or “trustless” consensus.
New ideas and new forms of distributed cooperative social and economic order are proposed.

This paper is organized as follows.
In Section \ref{sec:2:MainDirs}, we introduce the main 
directions 
and we show that there exists more than one point of view about the role of privacy.
In Section \ref{sec:3:Exchanges}, we elaborate on the central role played by crypto exchanges and discuss related risks.
In Section \ref{sec:4:ThreatCenter}, we discuss the major emerging risks at the center stage of crypto currency.
In Section \ref{sec:5:FromLawE}, we look at the concerns emanating from the law enforcement sector.
In Section \ref{sec:6:CryptoChoices}, we investigate the dangers of open source
and explain how much engineers are responsible for the crime and theft landscape,
which is also simply part of the DNA of our systems. 
In Section \ref{sec:7:PrivEnha}, we explain how cryptographers have been trying to improve the privacy of the original bitcoin in the last 10 years.
In Section \ref{sec:8:MythAnon}, we address certain misconceptions,
which exaggerate the purpose and the type of anonymity we obtain with privacy-enhanced crypto currencies.
In Section \ref{sec:9:ChalLawE}, we look at how law enforcement and regulators can cope with new emerging challenges.
In Section \ref{sec:10Klaus:CrimesClassifyProport}, 
 we look at three main types of crime by volume of activity: thefts, scams and ransomware.
In Section \ref{sec:11:NewxCryptoCrimBattlefiled},
we look at the idea of a criminal battlefields.
In Section \ref{sec:12:NewEcoNeedsLegalProt}, we look at how the emerging crypto economy can be better protected rather than banned.
In Section \ref{sec:13:ReguCryptBusi}, we study how investors
and asset holders may need legal protection and crime policing.
Then comes our conclusion section with some public policy recommendations.

\section{
{Main Directions and Historical Development of Crypto Currency}}
\label{sec:2:MainDirs}

\subsection{Legal and Technical Qualifications of Crypto Currency}
\label{sec:2:MainDirs:LegQual}

There exist various legal, regulatory and technical qualifications of crypto currency systems.
These qualifications are {\bf} not exclusive.
While courts of law or the United States CFTC recognized bitcoin as a “currency”,
the SEC has fined many companies for improperly issuing a “security”, the police have been chasing bitcoins related to major crimes
and recognizes that crypto currency poses unique challenges around increased privacy.
Then, national tax authorities e.g. IRS decided to consider bitcoin as “property”,
imposing truly excessive transaction reporting obligations and discouraging mainstream adoption.

This list is not exhaustive. For example, almost no one is calling bitcoin a lottery,
even though the current proof of work (POW) mechanism creates a winner every 10 minutes, see \cite{cg34},
and the winner is systematically awarded a substantial amount of 6.25 freshly created bitcoins.
Moreover we could also claim that bitcoin is a voting protocol.
In fact it is a self-governing financial cooperative:
the entities making the winning blocks are able 
to co-decide which transactions are accepted 
and also which version of the protocol and software is dominant.
This aspect of self-governance is substantially more developed in Ethereum and other blockchain systems
where the smart contract functionality seems to be the main one.
Most crypto currency systems are essentially also tools of "disintermediation":
they provide secure methods of holding
and controlling property and financial assets
in "direct" ways, without trusted intermediaries, banks, notaries etc.
This is achieved with help of cryptographic keys stored on secure devices controlled by users.
Likewise, they provide ways of controlling assets in completely new or decentralized ways.
Blockcahin and crypto currency systems are an ongoing challenge to our ideas.
They force us to revisit our ideas about questions such as
how the focus of crime policing should change with new technology.
We probably need to think twice about what is socially desirable to do in a market economy,
or 
how financial services regulation in the 21st century should look like.

\subsection{Waves of Disruption}
\label{sec:2:MainDirs:WavesFunctions}

The story begins with users who needed bitcoins for their para-banking ability: to store and transfer money.
An anarchic system initially, operating without too much consideration for national boundaries and regulations.
Such users will be, for example, families who send their children to study abroad
or small businesses which operate across borders.
All this is however far from being the primary function of bitcoin as of 2021.
Bitcoin has yet failed so far to become a champion of small payments
or to be adopted in the Internet retail commerce.
In 2020-21 we finally have the lightning network, a major enhancement in this direction.
Yet electronic commerce has not changed yet,
it is working primarily with Visa and Mastercard, as it did for decades.
Nevertheless a very substantial disruption is here.
We see it in particular in the world of investment.

If the first somewhat primary function of bitcoin is trans-border money transfer, the second would be the store of value.
Bitcoin has for more than a decade managed to establish itself as a reliable store of value
with a peculiar ultra deflationary monetary policy  \cite{cg29,cg34}.
Since late 2014, there have been very substantial press blogosphere and mass-media coverage related
to the crypto currencies and how they disrupt the world of banks and businesses.
For example, in 2019, twenty years after the creation of the Euro single currency,
Google search showed more searchers about bitcoin than about Euro.
This is happening, even though we buy our food in Euro 100 $\%$ of the time,
and 0 $\%$ of our house loans are denominated in bitcoins.
As of 2021, bitcoin is and remains a curiosity from the mainstream economy point of view.
Half-way between offshore para-banking, it is also a super-famous club of cyber-punk innovators and disruptors.
More recently we hear about institutional adoption and the ability of bitcoin to go up faster than inflation,
while governments are creating excessive quantities of fresh money in order to alleviate the pandemic-driven recession.
Investors have eventually diversified their portfolios and bitcoin became a mainstream financial asset in the recent years
and hundreds of millions of people own bitcoins, cf. \cite{CryptoComAdoptReport0721}.

It is totally incorrect to believe that crypto is just an investment bubble.
It is way more.
If it is a ``bubble'', it involves an incredibly large amount of coder, developer, R$\&$D
and cryptographic engineering activity, open source software.
We have a plethora of hardware and business infrastructure projects developed by expert firms.
Bitcoin adoption has increased by 880 \% in 2021,
according to Chainalysis \cite{Ad880Pc}.
It also involves a very substantial amount of academic research activity, which is not less significant than some of to the very major
intellectual endeavours of the human civilisation, such as sending a man into space, or tackling the Covid pandemics.
It is also a  ``bubble'' which leads to shortage of certain types of GPU computing hardware, as much as bitcoins are in short supply also.
Hackers, cryptographers and coders have created a monumental new technology development movement, which is self-funded.
It is somewhat funded precisely, by inflated investment expectations... and the unreasonable crazy ride which it offers to investors.
Crypto economy is poorly integrated with mainstream economy,
yet it has created a huge economic and startup company activity, which is unprecedented.

Bitcoin is a brand, a cultural movement, and it is religious at times.
Bitcoin maximalists believe that bitcoin prevails no matter what.
Bitcoin is an epic story, at the crossroads of “geek mythology” and libertarian anarchist cyberpunk ideas.
Blockchains are distributed electronic ledgers run with peer-to-peer cooperative consensus.
They allow keeping money outside of the presumably “evil” establishment and banks.
As a result, law enforcement is not facilitated.
It is more difficult, for example, to seize assets or to undo a transfer.
Beneficiaries can be anonymous, transaction flows can be obfuscated,
and income can be possibly concealed for taxation purposes.

\subsection{Privacy - A Two Sided Coin}
\label{sec:3:A:PrivTwoS}

There is a substantial amount of ambiguity about 
whether privacy is socially desirable, and to what extent privacy enhancing 
methods and techniques are legal or ethical. 
Privacy, is a tool of choice which can be used to obscure your financial history
for wrong reasons or an illegitimate purpose.
However there are also many strong arguments in favor of privacy. 

Privacy is a basic and constitutional human right.
People have a legitimate desire to escape the prying eyes and mass surveillance.
The Article 8 of the European Convention on Human Rights (ECHR) establishes
the Right to Privacy.
Initially this seems to have a limited scope,
with focus on family life and protection of private correspondence
and relationships with other human beings.
However, many courts and lawyers give it a much broader interpretation,
like to be free from “unlawful” indiscriminate searches
done at a large scale.
A recent ruling of the European Court of Justice CJEU of 6 August 2020 makes
mass surveillance programs run by the UK, French and Belgian governments simply illegal, cf. \cite{cg46}.

There are “public benefits” from increased anonymity, 
for example we have reasonable expectations not to reveal to your spouse, 
your employer or neighbor which political or social cause we support financially \cite{cg35}, 
for example with political donations in bitcoins. 
Lack of privacy in financial transactions greatly {\bf exposes} individuals 
to be easy targets for numerous serious crimes such as “identity theft, 
mugging, kidnapping and blackmailing” \cite{cg35}. 
For example, we consider investment scams which account for some 54 $\%$ of bitcoin criminal activity.
Here we observe that privacy and law enforcement are also at stake.
The more all sorts of data items pertaining activity are exposed, 
collected, stored, used, re-used, compromised, shared or sold, 
the more criminals can operate sophisticated targeted scams. 
Or they will simply succeed in hacking into our accounts.
A nice example of how the law can positively impact cybersecurity 
is Texas law HB 3222 from 2009 which forces businesses to comply with 
payemnt card industry security standards which in particular forbid storing certain 
items such as 3 digits printed at the back of credit card, 
cf. PCI DSS section, cf. slide 99 in \cite{cg12}.
We see that orderly and legal business activity also requires privacy. 
Data collection and storage needs some form of regulation. 
Bruce Schneier has as early as 2005 advocated that we need to ``reduce'' and ``limit'' 
data collection and ``how it can be used'', see slide 92 in \cite{cg12}. 
In general large-scale indiscriminate bulk data collection, or retention, 
or negligence when information is leaked out, is against the law in most jurisdictions. 
Large companies have been fined hundreds of millions of dollars or euros, 
for failing to secure personal information  under the new GDPR regulations in the EU, 
or older laws such as the U.S. GLBA regulations \cite{cg35}.

\subsection{Privacy vs. Law Enforcement - A Balancing Act} 
\label{sec:3:B:PrivBalanced} 


Privacy is not only a political and legal question. 
It is also a technical question where we can try to achieve a certain equilibrium. 
Privacy can help to evade taxes or circumvent business or financial sector regulations. 
However, poor privacy of financial transactions is precisely
a key factor which makes it easier for hackers and thieves to operate. 
If we have too much privacy, criminal activity is facilitated.
Then also if we have too little privacy, crime is also facilitated.
In order to achieve a reasonable balance between privacy and law enforcement
we need to consider 
several key questions.

Even though blockchains, in theory, 
are self-policing for certain questions such as theft of funds in transit, 
they are not very good on protecting customer privacy. 
Initially, the whole idea of ``public'' ledgers is corrosive for our privacy: 
they expose our financial and personal data. 
Then we add privacy to these systems, 
for example with some cryptographic techniques, or simply through mixing and obfuscation. 
At the end privacy becomes potentially too strong or so it seems, cf. Section \ref{sec:8:MythAnon}, 
and criminal activity is potentially facilitated. 
A question is not quite to see who wins: the policemen of the thieves. 
It is a classical problem in technology and in law enforcement. 
Both sides follow a certain curve: 
security and safety improves, and criminals and thieves also evolve. 
The question is really about achieving a better more detailed understanding 
of how technology actually impacts our lives. 

We need to put things in proportion. 
Expert studies 
show that the criminal segment is small in bitcoin, 
or small in relative terms; 
it is about 0.17 $\%$ on average and yet smaller in many jurisdictions, 
for example it is only 0.06 $\%$ in Germany, cf. \cite{cg24}. 
Very similar statistic were observed in credit card payments in the last two decades: 
the fraud is 0.15 $\%$ typically and significantly smaller in well-policed jurisdictions \cite{cg12}. 
Yet nobody says that credit cards are best friends of criminals. 
In both cases, we see that the legit payment activity is substantially larger. 
To summarize, criminal activity is small in bitcoin and elsewhere, 
and 99.9 $\%$ of crypto currency activity is legitimate trading or commercial 
activity. This part arguably should get more attention than the relatively small illicit 
activity.

\subsection{Investment: Major Not Minor}
\label{sec:3:B:InvsMAJR}

A major angle to consider is gambling and investment questions, which are very hard to tell apart.
On the one hand, we have a legitimate desire of the public to support and invest in crypto currency startups, as such businesses
which are not listed on the stock market.
On the other hand, we have the gambling question.
In the eyes of millions of small investors, crypto currency markets are
an offshore-operating casino with less stringent regulations.
We have the popular “rich-kid” dream of making money through the crazy appreciation of bitcoin and other crypto assets \cite{cg01}.
Here bitcoin competes with the stock market, Internet casinos, and amateur Forex trading to offer some thrills and hopes to millions of investors, frequently millennials.

According to \cite{cg39} as many as 11 $\%$ of Americans own bitcoins.
Every single day, 1 million bitcoin wallets are active.
There are more than 25 million bitcoin traders worldwide and “speculating is the most popular use of bitcoin” cf. \cite{cg39}.
In addition, according to a detailed industry report,
the rate at which the number of users of crypto currency
has been growing faster than exponentially.
In just four months between March 2021 and July 2021, the global number of crypto currency
users has doubled from 100 to 200 million approximately,
cf. page 9 in \cite{CryptoComAdoptReport0721}.

\newpage 

\section{
{ Main Business Players - Crypto Exchanges}}
\label{sec:3:Exchanges}

In theory, Satoshi have imagined a utopian peer-to-peer economy where we exchange coins against goods and services.
In reality we already have a way more sophisticated economy, with intangible assets and major extra features:
debt and leverage, network effects, business competition, and non-linear behaviour.
The central players where the old and new worlds meet are the crypto exchanges.

\subsection{Geography of Crypto Exchanges}
\label{sec:3:A:ExchGeog}

It is not true that bitcoin operates outside of the scope of law and order.
In fact, as much as 20 $\%$ of all bitcoins sent to exchanges globally are sent to exchanges based inside the U.S.
At the same time, 59 $\%$
of the volume goes to the so-called ‘high-risk’ exchanges \cite{cg49}.
These exchanges are notorious
for being lax on KYC (Know Your Curstomer)
and AML (Anti-Money Laundering).
At the same time, they are also simply small independent offshore companies where customers are at risk of losing everything.
It is important to see that these companies themselves can be scams,
and they can just disappear tomorrow with funds belonging to naïve crypto currency holders.
Users do however use them on a massive scale.
Following \cite{cg49} about 24 $\%$ of volume from US-based exchanges goes to these ‘high-risk’ exchanges,
while for Russian exchanges this percentage goes up to 69 $\%$.

\subsection{On Neeed for Investor Privacy}
\label{sec:3:B:NeedPriv}

We have seen that for all sorts of reasons, which could include tax evasion, individuals use and trust riskier crypto exchanges with their coins.
A major reason for that is that these companies ask less questions
and it is easier to pass various verification levels required and to remain anonymous.
We hypothesize that it is NOT in order to escape government surveillance that crypto currency privacy
is aiming at. 
Regardless of whether we like it or not,
amateur trading is a big industry and investors need to be protected.
First of all against losing all they have in investment scams and magical get-rich-quick schemes.
Some 16 billion dollars were lost in some 132 major crypto currency scams 
cf. \cite{cg50} and \cite{Scam16G}.

In fact, there is a balancing act to make:
tax evasion deprives governments from some revenue,
but there are substantially larger losses at sight.
Would these amateur investors lose all their savings,
governments worldwide will be facing larger bills for healthcare,
social housing etc., over extended periods of time.
Governments are the ultimate {\bf risk takers} for amateur investors,
for example assuring minimal income for pensioners robbed from their savings,
which normally should allow them to retire comfortably.
Thus, very clearly, investor privacy can indirectly {\bf protect the governments},
against 
future {\bf liabilities} at an astronomical scale.
This is given our aging population 
and the immense popularity of amateur investing.
We need to defend small investors against private sector predatory surveillance.
Activities meant to deprive investors of their fundamental rights
such as freedom to buy and sell with some privacy and without being ``gamed'' or manipulated.
We have the right to participate in a fair and open market economy.
One which is not rigged by current omni-scient monopolistic
big data and AI players,
which tend to abuse their dominant position.

\subsection{On the Neeed to Police the Investment Platforms}
\label{sec:3:C:ExchPoli}

Crime and law enforcement occur when people avoid paying taxes.
Larger crimes occur when companies do help tax evaders and organise the tax evasion at a large scale.
In November 2020 the U.S. Department of Justice pressed criminal charges for BitMex,
an exchange based in the Seychelles for "serving U.S.-based customers" making it very clear that
U.S. law will be applied in such cases \cite{cg42}.
Law enforcement also occurs when SEC fines companies that raised money in an ICO,
selling a token which is considered to be a security.
Several businesses have been instructed to return vast sums of money to investors
by the policeman of the financial markets, the powerful SEC.

Market manipulation is also criminal.
Policing this crime can arguably involve increasing citizen security against predatory banks.
Banks have been fined huge sums in billions of dollars in the last decade.
One research question is the one of adversarial trading and front-running cf. \cite{cg42}.
A key problem is that blockchains has so far focused on how to solve consensus problem,
which is mainly about blocks that are eventually approved and final.
Now there is {\bf a larger space} for adversarial traders:
the question of incentives and selection of transactions accepted within one block \cite{cg42}.
Then we have the questions of adversarial AI:
predicting the behavior of humans, in order to play games against users in the marketplace.
This does not seem illegal per se,
but in fact it will clearly be unethical and fraudulent,
if you pay a bank or the exchange some fees or subscription,
in order to advise you about investment opportunities.

\section{
{Major Risks and Threats at the Center}}
\label{sec:4:ThreatCenter}

It is estimated that hundreds millions of dollars per year are made by criminals in extortion of ransom payments,
which have plagued countless businesses worldwide, small and large \cite{cg09} and Section \ref{sec:10:A:Ransom}.
Yet the illicit activity amounts to just 0.1 $\%$ of crypto deposits typically, and this figure went down by 60 $\%$ in two years \cite{cg24}.
Millions of people go to crypto currency markets not in order to acquire illegal goods, but rather to simply acquire bitcoins,
which for most users are perfectly legal to own. Then we sell them at profit or at a loss.
Individuals can also mine bitcoins at home, see Section \ref{sec:6:B:SHA256Centr}.

\subsection{Is Bitcoin Itself a Financial Scam?}
\label{sec:4:A:BTCitselfScam}

For more than 10 years, bitcoin has been traded.
The value if bitcoin did not go to zero.
Bitcooin holds hundreds of billions of dollars in funds belonging to hundreds of millions of users
\cite{CryptoComAdoptReport0721}.
The public perception of the evils and dangers of bitcoin has been distorted by the idea,
that certain incumbents in the financial industry, and some governments alike, would like to
make something like bitcoin or blockchain illegal. However.
The interests of small investors and the monopolistic established industry incumbents are not the same.
Here we present both points of view.

In some sense, bitcoin potentially IS an investment scam,
or we need to consider that it could be one, for example as a precautionary safety measure to protect investors.
It is a fact that banks are losing their deposits,
which go to a para-banking offshore space, with less regulations and no borders.
This makes laws harder to enforce.
The world of investment is extremely old-fashioned slow and rigid:
for example it is  very hard to buy and sell stocks from Asia for an investor who lives in Europe.
New blockchains, currently under development, promise to allow anyone to trade any asset freely.
We see that banks and stock exchanges are losing some of their sources of income and the competition is increasing.

This ongoing process of disrupting a multi trillion-dollar global financial industry
increases the risks due to bad choice of a financial partner or broker and
due to the new technology itself such as hacking.
It creates countless opportunities for negligence, misconduct, 
careless or “frivolous” behavior cf. \cite{cg27}
where a small individual investor will find it very difficult to know
which companies and which technical standards they can trust.
Investors can lose everything when hackers steal private keys.
Open-source software can be malicious \cite{cg26,cg27} or just insecure.
Current laws and regulations favor the incumbents, and maybe only very large companies can apply.
The financial industry remains highly regulated and is a subject of intense government oversight and scrutiny.
Law enforcement must simply {\bf expand} and regulate as much it can.
We need to extend the safety net, allow smaller business to operate, and protect individual investors.
Law enforcement needs to follow the investor money,
and contribute to defining how the crypto industry should be run as a whole.
We also need to forbid or allow and regulate more clearly.

As already explained, the criminal or illicit segment is very small essentially 0.1 $\%$ \cite{cg24}.
The next level will be “Joe the Plumber” stashing some cash for his retirement
at an offshore crypto exchange, which income possibly he forgot to declare with IRS,
which raises serious questions \cite{cg40}.
Here the source of funds is not illegal itself.
It is not illegal either for Joe to avoid his bank,
which pays very little dividend or/and charges him large commissions on his retirement investment.
We live in a free market economy, and it is not a crime at all for anyone, to seek a place to deposit money,
where they can generate a better income.
When local economy and or the established investment providers suffer some setbacks, investors do and will go to exotic places.
One day it was gold, or some tech companies, today bitcoin is a popular
asset for 1 million amateur traders each day \cite{cg39} worried about inflation or maybe a Covid-induced recession.
The question is now simply, that startups and offshore firms which do not have a banking license, and
yet are offering financial services to Joe. Potentially they are incompetent, cheating,
or just too easygoing optimistic startups relying on untested cryptography and software,
or neglecting public safety for the sake of new exciting functionalities.
The risks for small investors are simply enormous, and banks are regulated for a reason.
The main idea is that small or amateur investors deserve accurate information
and some protection 
against fraud, misrepresentation, or more ordinary "preventable technical incidents".

\section{
{Anarchy Meets Law Enforcement}}
\label{sec:5:FromLawE}

Our sources of information are academic research, the chaotic world of Internet forums and collaborative platforms,
the abundant press coverage of crypto industry, and interviews with law enforcement professionals.
Throughout the duration (2014-2017) of the European Commission FP7 PRIME Project
(Preventing, Interdicting and Mitigating Extremist Events)
we have interviewed 132 front-line investigators
and law enforcement officers with security merit from
Australia, Austria, Canada, Germany, Georgia, Hungary, India, Italy, Japan, Mexico,
New Zealand, Poland, Portugal, Spain and United States, cf. \cite{cg03}.
We also asked our interlocutors for their opinion on the abuse of modern technologies by violent extremists.
If this paper contains certain statements which are a matter of opinion, this will not be necessarily ours:
they come from interviews with law enforcement professionals.

We found that the awareness of blockchain technology among our target audience (specialists with a wide array of law enforcement backgrounds:
police, border guard, intelligence, counterintelligence, military, government and special services) is very low and vague.
In 2017 only a small number (less than 10 $\%$) of our interlocutors understood the significance of blockchain
and appreciated the potential for the criminal abuse of technology.
It needs to be stressed that these officers were experts in financial analysis
and represented top-level law enforcement institutions in their respective countries.
Those specialists indicated that the financial analysis techniques can and are used
to {\bf retroactively } study the cash flows and money transfers (including crypto currency such as bitcoin)
in the cases of known lone-actor extremists,
in order to search for patterns that could be later applied
to investigate radicalized individuals \cite{cg04}.
However this approach is potentially ineffective and inefficient for the purpose of regulating
ordinary crypto market activity. It does not prevent thefts and scams, and does not help
customers to recognise reliable providers and reliable investment opportunities.
On the contrary. Many people are lured by exotic places where maybe the taxman or their spouse are no longer watching,
and take substantial risks for no reason other than, a quite naive expectation of better privacy.
It is naive, because unregulated offshore service providers are in fact MORE likely to sell data to criminals than reputable crypto companies.
Arguably we need more "legit" providers of financial privacy.
A UK government report explains that Suspicious Activity Reports (SARs) do not help a lot \cite{cg53}.
They are filed in "huge numbers" and are typically just personal living expenses.
The safeguards which help to catch terrorists or to tackle organized crime,
are potentially 
not that useful in regulating a broader para-banking activity.
A broader population of legitimate participants in these markets lacks
sufficient protection against poor security standards or negligent service providers.

\subsection{Risk Awareness, Blockchain Education, Trends and Geography}
\label{sec:5:A:EduBlockchainRisk}

When consumer protection and technology fail,
or when regulating or banning does not help,
or when policing does not reduce crime and fraud,
we need to work on public awareness and education.
First of all, we should not underestimate the educational benefits of crypto currency markets.
Instead of unemployed youngsters depressed by 
anti-capitalist propaganda and gloom,
we have a game of building a new world order by geeks, coders and hackers.
A dream, of becoming a banker, creating money out of nothing, being a stakeholder participating in the economy of the future in active ways.
The term ``cryptographer’s dream'' is sometimes used \cite{cg25}.
A new form of cooperative decentralized and inclusive economy is possible.
Middle class people worldwide learn about new technology, maths and cryptography,
consult various blogs and websites, and develop a feeling of being autonomous players and exploring the world of alternative finance.
Coders and designers are empowered to build financial systems,
which work the way {\bf they} think is ethical and socially desirable.

Even though blockchains were invented more than 12 years ago \cite{cg02},
it is still -- in 2021 -- a new and emerging phenomenon and the knowledge about it is poor.
This is clearly indicated by the patterns in search queries 
indicated by Google Trends analytics.
We have seen a 10-fold increase in searches about the word ``blockchain'' in 2017,
then it goes down 3-fold but never reaches the pre-2017 levels, and again a 3-fold increase occurs at the end of 2020.
The curve has two sharp peaks which correspond exactly to moments when bitcoin price has peaked at the end of 2017 and again at the end of 2020.
The pattern is similar among the law enforcement and intelligence/security communities.
Google trends also show that peripheral or emerging countries do more searches than Western countries.
The ability of bitcoin to offer alternative international money transfer services in various countries
was just the starting point. 
Blockchains have now become an important piece of the liberal globalized English-speaking economy.
Among 78 countries Ghana and China are those,
where most searches about blockchain originate from, with a scores of 100 and 92/100.
Most EU countries and the United States are way more conservative:
they arrive at positions 30-50 out of 78 with scores between 12/100 to 20/100.
Another detailed research report found that Vietnam followed by India and Pakistan
lead the crypto currency adoption in 2021, \cite{Ad880Pc},
when adjusted w.r.t. the purchasing power per capita. 
We discover that new tech such as blockchain catalizes progressive pro-business
and pro-education forces in
emerging 
countries.
This is great news.

\newpage 
\section{
{On Critical Technical Choices}}
\label{sec:6:CryptoChoices}

\subsection{Trusting the Cryptographers}
\label{sec:6:A:TrustCry}

There is a level of disconnection between traditional crypto research and development of industry standards
(e.g. in telecom, smart cards, and at NIST, or in EU-funded R$\&$D projects) and the nascent crypto industry since 2009.
Developers in bitcoin and other crypto currencies have made many deliberate choices in their specifications
which are peculiar, risky or non-standard features.
This contrasts a lot with the dominant culture in crypto research, which have been for 50 years very cautious,
assuming that the attackers are more powerful than they really are, and engineering in truly paranoid ways.
Traditional cryptographers have been very cautious and always considered that when something
is very special it should be banned.

For example AES is the most important standardized encryption algorithm today.
For this reason alone it is also used by the bad guys, e.g. in ransomware \cite{RansomCECC20}.
When the AES was standardized in 2000,
there was some controversy around the special choice of S-box used,
with New Scientist title ``Cipher Crisis'' in 2003 \cite{NewScientist}.
Bitcoin and crypto currency developer communities have ignored many traditional rules of careful security engineering,
and are in general very easy-going: ship it first, tell users this payment system is experimental,
and hope that hackers and thieves do not have sufficient expertise to steal coins and exploit numerous vulnerabilities in existence.
Bitcoin does not use AES, it uses SHA256, RIPEMD-160 and ECDSA with secp256k1.
The last part is particularly controversial cf. \cite{cg29}.
In fact secp256k1 is not what other people in the industry use for securing Internet commerce with TLS (secp256r1 is used 90 $\%$ of the time).
The crypto currency community has developed its own set of {\bf incompatible} crypto standards.
Similar but not quite the same or not exactly the same, as found in more traditional security industry such as in smart cards
or in TLS. Bitcoin does not obey global de-facto crypto standards such as defined by the U.S. NIST,
rather it defines its own standards used primarily in crypto currencies and virtually nowhere else.
They differ from security protocols used by major companies which define the technical standards in
major operating systems and Internet commerce (Microsoft, Apple, Google etc.).

Many advanced cryptographic protocols specify further new elliptic curves with further very special properties.
For example, Ethereum, ZeroCash, Algorand and Chia network use a special curve called  BLS12\_381.
Internet Computer (a.k.a. DFINITY) uses also more exotic pairing-friendly curves such as BN254N, BN\_SNARK1, BN382M, BN462, see \cite{ExoticCurves}.
In general the more advanced is the crypto market functionality or privacy enhancement functionality we find in blockchain technology,
the more risky and untested is the cryptography used there, see \cite{cg28}
and how researchers are “cavalier about proposing untested cryptosystems”, see \cite{cg32} and \cite{cg33}.
Innovation requires some ``riskier than usual'' security assumptions, and this is normal.

However. New cryptography standards are made and broken each year.
History knows numerous examples of faulty cryptography standards, which has nevertheless dominated the market very heavily.
For example, the Data Encryption Standard (DES) was developed in the mid 1970s and is still used in 2021.
Hundreds of millions of people use DES every day,
when they use their Chip and Pin bank card, and also in today’s contactless payments.
This cryptographic algorithm is far from being perfect, 
see in particular page 6 in \cite{cg30}.
For example in the key scheduling in DES there "is no interaction across the two halves", 
plus countless other known problems with DES, cf. \cite{ShamDESB,cg31}.
Yet DES has thrived for 50 years and remains massively used in financial systems. 
More recently the cryptographic hash function SHA-256 is used to secure the integrity of
bitcoin transactions and to secure the chain of events in essentially all current blockchain systems.
If this cryptographic hash function was faulty, it would be extremely difficult to upgrade.

Current systems do not consider that cryptographic algorithms could become obsolete overnight,
and we would need to switch and use a different crypto algorithm.
This question of the ``strategic'' choice of the crypto hash function is
also related to feasibility of specific types of attacks and crimes, as we will see below.

\subsection{
How Specific Crypto Engineering Choices Influence Fraud and Crime}
\label{sec:6:B:SHA256Centr}

An interesting example to learn from, is a story of an electric scooter one of the authors purchased in 2020.
Then he visited a large bike store and found that no lock currently on the market had the dimensions and shape,
which would allow one to attach this scooter reliably. The market does not provide a solution.
As result the scooter was stolen 3 days after purchase.
In addition, it did not have at all any unique serial number which could be communicated to the police.
How does it relate to crypto currency market?
It is a question of {\bf market regulation}:
potentially certain models of electric scooters should be banned from the EU or US markets,
because they lack consumer protection or any protection against theft whatsoever.
It is a problem of industry standards, holistic crime prevention and reconciling government
and private sector industry regulation
when new technology is sold in the marketplace.

Going back to crypto currencies, there is no doubt that several technical features of bitcoin
facilitate specific types of crime, cf. last point in Benefits section page 9 in 
\cite{cg52}. Is it possible to claim that some {\bf features} of bitcoin could be banned?
Yes without any doubt, and the key problem is that bitcoin failed so far to become decentralized, on the contrary,
cf. p.17 in \cite{cg52}. 
One can argue that bitcoin developers could have, in early days, mandated a different hash function in bitcoin.
One, which could be mined with GPUs which are readily accessible in every computer store.
Sticking with current NIST crypto standard, which however is GPU-incompatible,
namely SHA-256 for mining, has immense consequences.
In some sense bitcoin mining is not a mainstream system, because bitcoin developers decided not to go this way.
When crypto miners started making specialist ASIC bitcoin mining equipment,
such devices were not sold in any of traditional trusted merchants of electronics.
They were just sold by mail/Internet order by extremely few small companies in remote countries.
As a result, there were multiple scams, and companies who failed to deliver these machines.
Thus, bitcoin mining has been monopolized by larger investors.
They were the only ones able not to fall victim of dishonest providers (a repeated game).
This is NOT possible for individuals, who just order one unit from a foreign manufacturer advertised on the Internet.
All this is a consequence of a questionable choice of a crypto hashing algorithm, which has led to
a large degree of 
centralization of bitcoin mining.
Criminals have sold non-existing bitcoin miners to naïve investors,
and a key problem was that there was no reliable supply chain whatsoever for small individual buyers, cf. \cite{cg34}.
There are additional problems which also lead to centralization.
The reward from one block is very high which leads to mining with pools, which are a centralizing force
and which have heavily distorted the Satoshi original dream of a peer to peer anarchic cooperative system.
Then, the incredibly large electricity consumption by bitcoin creates serious problems.
One problem is that for many years critics claimed that bitcoin is a de fact Chinese crypto currency:
as it is controlled by a few large Chinese miners\footnote{This has started changing in 2018-2021
when Chinese authorities have taken numerous decisive steps in order to ban or discourage bitcoin mining.}.
Why this matters? It is easy to see that miners can influence the future of bitcoin,
such as adoption of upgrades. They can also potentially censor individual bitcoin transactions,
or simply privilege transactions originating from China.

Looking beyond China, there is a group of countries which officially are recognized as supporters of terrorism.
These countries are then logically subject to international sanctions which attempt to prevent them from selling their oil and gas.
It follows logically that they may use bitcoin in order to re-sell their energy production surplus.
We see that bitcoin hash function is an unfortunate choice upheld by generations of developers in 2009-2021.
It makes that terrorists, or just ordinary authoritarian regimes,
are likely to mine bitcoins, which makes wider adoption of bitcoin problematic.
However, due to the choice of the hash function, this was never a problem with Ethereum,
which can be mined with every GPU available in most computer stores worldwide.
More recently in 2021 we have seen GPUs which are decidated for gamers, and where crypto mining is disabled
and its performance is degraded.
We have also seen the emergence of Chia crypto currency which make extensive use of computer hard drives.
We see that a bad crypto hash function choice can be toxic (in bitcoin) and a good choice can be beneficial to some segments of the computer industry.

This consideration was here for a long time, and countless other crypto currencies
had in contrast mandated a hash function, which can be mined even with ordinary PCs.
However, this choice can backfire on the crime front.
There are hackers who break into computers of millions of people, in order to run networks of zombie machines,
which are sold on the black market in order to be rented to other criminals. 
We also have scammers and spammer activity.
It appears that the most frequently mined crypto currency by such zombie networks is Monero.
This is precisely because, it can be mined on more or less any PC.
This has serious consequences \cite{cg43}.

There is yet another side to this debate about the choice of a hash function.
GPU mining with any PC is what we need if we want privacy-enhanced tech
to protect us against mass surveillance, or just say burglars,
who will use the Internet to find out that when we are not at home.
Privacy can only improve, if millions of users can effectively conceal
our precious transaction data in a larger cloud of user activity.
Thus, it is preferable that a privacy enabling coin can be mined on any device.
With more users, we obtain better privacy for our financial transactions,
and credentials used to access them (password, cookies, etc).
Privacy is a basic right \cite{cg35}, and also a necessity, in order to prevent various forms of crime and theft,
such as targeted phishing attacks against banks and trading platforms,
or targeted kidnapping cf. Section \ref{sec:10:C:Kindap}.

\section{
{Privacy Enhanced Payments}}
\label{sec:7:PrivEnha}

It is important to see what kind of privacy enhancing techniques are available today in blockchain and bitcoin space.
For bitcoin we refer to Section V in \cite{cg52}, dated 2017.
A high level overview of various bitcoin anonymization techniques and main privacy altcoins
can be found inside \cite{cg52,cg35,cg28} and many others.
In this section we also briefly review these techniques.
Our approach is to look at the history of ideas and crypto technology adoption.
The legality of privacy-enhanced coins and AML regulations are covered in \cite{cg35}.

Bitcoin was a first very approximative creation of Satoshi Nakamoto \cite{cg02}.
It is easy to see that later bitcoin is NOT at all as Satoshi has imagined it \cite{cg29}.
Nakamoto describes a system with a large number of participants where everybody is mining,
connecting to peers and engaging in transactions.
He did not anticipate that these 3 groups of people are going to become almost entirely disjoint,
and that we will have huge for-profit mining farms using huge amount of energy,
millions of wallets connecting through trusted gateways and vast amounts of activity,
which are not at all recorded on the bitcoin blockchain.
Moreover the number of people who actually run the P2P network remains very low,
between 5,000 and 11,500 in recent 5 years, source: www.bitdes.io.
Bitcoin has many serious problems \cite{cg29,cg52} and this is precisely the challenge for the next generations of crypto engineers.
It has led to creation of hundreds of “altcoins” with distinct technical features trying to fix some of these problems like cost, speed \cite{cg36},
number of transactions per second, resistance to attacks by powerful entities, etc.
Among others, anonymity is not as good as it initially seemed.

Initially, bitcoins are attributions of certain amounts of coins to public keys.
A possibility inherent to the so-called public key cryptography.
The owner of the private key can spend the coins. We have a pair of cryptographic keys which match each other.
To a large extent bitcoin is an evolution of the French invention of the Chip-and-Pin bank card.
A card has a private cryptographic key,
which allows to certify that one specific transaction was approved by the owner;
the date the amount and other details are digitally signed.
This is in fact done more than once, with DES and with a public key algorithm,
and at several stages ever before the transaction is completed \cite{cg12}.
Then, with bitcoin the card is no longer issued by a bank but by a community of peers,
which creates their own coins and attribute them to each other.
Bitcoin is simply the next logical step in an evolution of commerce underpinned by public key cryptography.
The reliance on security and secrecy of cryptographic keys increases tremendously,
and bitcoins which are not stored on exchanges are typically stored on secure hardware
devices. These devices are evolutions of a Chip and Pin banking smart card.
More than 1 million Ledger Nano devices have been sold in the last 5 years (source: ledger.com).

Bitcoin can already offer pseudonymity:
a user can use many different public keys, which do not identify him easily.
Coins from different sources can be mixed at will.
This works, if we do one transaction, however the more activity we do, the more information is leaked to attackers.
We need to use more advanced cryptography to do better.
Interestingly, almost all known ways to enhance privacy are simply and directly
linked to more advanced types of public key cryptography.
For example, with stealth address techniques cf. \cite{cg29},
coins can be sent to a recipient and we are not able to link this recipient to the address the recipient himself advertises or makes public.
It is simply a more private form of PK cryptography with additional tricks,
where it is the sender not the receiver, who creates ephemeral bitcoin public keys used to receive incoming payments. 
This is used in Monero crypto currency, one of the most frequently cited as used or mined by criminals \cite{cg40}.
As a result, Monero is now an object of sophisticated de-anonymization efforts for law enforcement purposes \cite{cg24}.

Then ring signatures, also in Monero, are anonymous decentralized group signatures. A signer can be any of say 5 signers selected in one transaction. The power of this technique is that the other 4 signers do not need to cooperate.
Anyone is by default involved or can participate in anonymous transactions without their approval,
enhancing the privacy for everyone.
Then Zero-Knowledge proofs can be used to create methods of generating transactions
with a far larger degree of ambiguity about who the signer is.
It is a type of advanced public key cryptography, which is notoriously difficult to make.

We see that crypto currency privacy innovation is primarily technical.
Possibly and arguably the market and the economy does not yet require such high levels of privacy, however tomorrow it will.
Cybersecurity is a never-ending quest of enhancing privacy and anonymity and security against fraud and theft
and standards are always improved and extended, in order to eventually do more than necessary,
and they rarely get simpler. 
There is no end to this process.
We need to understand that advanced cryptographic techniques are primarily
here in order to police bad actors who cheat or steal the funds in various ways.
They prevent crime and fraud as much as they might help criminals.
When anonymity weakens the certitudes, new types of thefts and attacks can emerge.
There is a genuine expectation to enhance privacy, with intention to build better, fairer, and more secure payments systems.
However, unquestionably there exist evil geniuses of crime, enabled by new opportunities such as with ransomware and investment scams.
Then we have rogue engineers with dubious goals, or the fact that cryptography and computer systems fail most of the time and
good security is very hard to achieve.
Which brings again the question of the quality of the source code \cite{cg26}
and of regulating or just better organizing (self-regulation)
the chaotic payment and investment innovation world.

\section{
{The Myth of Crypto Anonymity}}
\label{sec:8:MythAnon}

A popular misconception in the existing literature press and blogosphere
is the alleged full-anonymity (or at least pseudonymity) of different crypto currencies.
In effect, the services responsible for maintaining public security and safety fall victim to the misconceptions
and fear-mongering related to blockchain and crypto currencies,
which is further amplified by the pressures of the governmental agencies supervising the work of law enforcement,
with some countries going as far as endorsing the idea of making blockchain ``illegal''
or supporting the all-inclusive state control over it \cite{cg05}.
In recent years we had multiple events of crypto currency exchanges
in regulated territories such as US or South Korea officially {\bf de-listing} several privacy coins due to regulatory pressure.
It seems that they have been asked by law enforcement agencies to produce
excessive quantities of SARs, 
which imply a certain cost burden, and yet are not the useful in crime prevention, cf. pages 58 and 66 in \cite{cg53}.
According to the aforementioned interviews completed throughout the duration of the European Commission PRIME Project,
as well as the further studies performed during the preparation of this paper,
the most serious concern among law enforcement community
is the potential use of the crypto currencies as a vehicle for the financing of terrorism \cite{cg03,cg04,cg53}
and specifically, transacting with countries subject to international sanctions.
This is confirmed by the existing (and scarce) literature, where the authors explicitly
refer to blockchain or bitcoin (or maybe Monero and similar privacy coins \cite{cg40})
as an obvious method of providing funds to extremist and terrorist organizations and radicalized individuals
(so-called lone-actor terrorists or “lone wolves”) \cite{cg06}, \cite{cg07}.
Such observations are amplified by the (exaggerated) popular perception of crypto currency being again completely anonymous
\cite{cg08} with some authors claiming that the payments and transfer completed with bitcoin are untraceable \cite{cg09,cg44}.
The claim becomes more serious when security researchers indicate the existence of techniques such as crypto currency “tumblers” or “mixers”, with specific examples of existing services which combine transfers that are happening at roughly the same time, and/or then re-route them to the final destinations \cite{cg10}. In general, tumbling techniques obfuscate the provenance, possession, and movement of crypto currencies \cite{cg11}.
This seems extremely strong but is not, due to the fact that the communications are in general monitored,
and their meta-data are retained and stored, which is enshrined in legislation in most jurisdictions worldwide.
Until recently the law enforcement community has a very simplistic view of these issues
reflected by a very limited vocabulary and simplistic answers (such as: “forbid crypto currency”).
Actual criminal or terrorists operations are very few, and operate with cash primarily cf. 
\cite{cg53}. Crypto currencies are used primarily to solicit donations.
According to Chapter 8 in \cite{cg53} the role of crypto assets in terrorism financing
has increased in 2020 and the score was changed from low to medium.

It is not productive to just claim pretended anonymity or lack thereof depending on who is talking.
Inevitably a specialist view is emerging with a more detailed vocabulary.
In recent years, criminal investigations have just shifted to specialized high-tech entities,
which have developed the necessary expertise and tools \cite{cg24}.
From an applied cryptography point of view, nothing is completely new and there is a long game where payment cryptography and security develop in many steps \cite{cg12}. Even Monero was discovered to be traceable in open academic research, which methods are now implemented by specialist firms \cite{cg24}.
The main point is that privacy is temporary, and can be disappear if additional future events take place, or if additional countermeasures against anonymity are in place, or if additional data are collected and appropriate actions are taken by certain entities.
When some academics are trying very hard to enhance anonymity \cite{cg44},
others are wiser showing that privacy is impossible to achieve in general \cite{cg45}.

We need to get the basic facts right. The primary reason to use cryptography is to prevent theft and enable more business to make remote payments easier.
Cryptography makes owning and controlling assets easier.
It also creates new types of thefts which have never existed before, cf. \cite{cg13}, \cite{cg14}
however this is now within limits of cryptographic engineering which is fragile and knows many points of failure.
In particular problems at operation due to human factors cf. \cite{cg13}, \cite{cg14} and unexpected additional events which are deliberately engineered \cite{cg45}. A mix of cryptography and secure hardware devices with anti-cloning protections is nevertheless an undeniable helper at the heart of our ``Chip and Pin'' bank cards and today’s bitcoin wallets. Users can physically control payments and authorize transactions themselves and this crypto + secure hardware revolution was inevitable and was started by the banks themselves as early as in the 1980s \cite{cg12}.
A secure hardware device \cite{cg12} with a private cryptographic key has three primary functions. One is to prove authenticity and to authorize payments or transfers remotely, over large distances while minimizing disclosures. Another is to prevent theft, for example through tamper-resistance and user authentication with a PIN.
The third is an anti-clone functionality, which is very strong with digital signatures, because the private key can be used but not compromised, and it will never even leave the secure element inside the smart card. No one can obtain or extract it, and the bank card is already a trustless payment system because users need only to trust themselves, so that they no longer need to trust the bank which does not know the secret keys. Inevitably, payment industry embraced this technology, as on all the three accounts (remote acceptance, anti-cloning, anti-theft), it is vastly superior to any payment technology, which was developed earlier, for example with paper money and bank notes. New forms of electronic money and new payment technologies are driving old solutions out of business (potentially) where cryptography is a crucial ingredient. There is no end in making financial systems more robust and less prone to subtle insider threats and attacks, and such systems naturally distrust every single transaction participant, including the banks and their computer systems themselves.
Interestingly, anonymity was yet totally absent. It arrives today, three decades later, and it comes in many different flavors, such as receiver anonymity, sender anonymity or un-linkability, trace-ability for transfers, etc. Now it crucial to see that anonymity is not something that cryptographers produce, or something they are good at (!). Cryptography can traditionally encrypt data, which is not really done in blockchain where ledgers are highly transparent, designed for broader verifiability. Instead, cryptography is meant to authenticate payments and transactions through hashing and digital signatures. This part of cryptography is very solid, and the primary reason to use cryptography is that it solves the theft and remote authorization problems extremely well, as no one can steal bitcoins by forging a digital signature. This strong ability is the primary reason why bitcoin does not need a bank or a notary anymore, and anyone can be their own bank if they wish to.
However, it is a misconception or it is rather naïve to believe that cryptography is here or even that it can at all solve personal and financial privacy problems. We have a broad spectrum of “privacy enhancing technologies” and a lot of innovation such as Zero-Knowledge proofs, all of which however are in fact imperfect.
Yet privacy is potentially impossible to achieve in general \cite{cg45}.
All we need is to apply current law enforcement and AML rules \cite{cg35}.
Privacy does not really exist, when the law enforcement has the ability to monitor all our communications metadata,
which is a strong legally enforced capability in most jurisdictions.
There is only a handful of well-identified geographical hotspots for illicit activity \cite{cg24}.
Cybersecurity is in general an endless game between attackers and defenders, and even if cryptography makes unbreakable locks,
so that bitcoins cannot be altered or stolen even by very powerful entities,
modern cryptography is yet weak and poorly suited to solve or address our privacy preoccupations. 
In fact potentially no technology can solve these problems in a satisfactory way \cite{cg45}.
The law enforcement and security services simply need to adapt and pay more attention to technology (and cryptography),
which works on both sides, helping attackers and defenders alike.

\section{
{Challenges for Law Enforcement}}
\label{sec:9:ChalLawE}

A more balanced assessment comes from the United Kingdom National Risk Assessment of Money Laundering and Terrorist Financing \cite{cg15}
which -- in 2015 -- rated digital currencies as a low risk, saying that (at the time) the criminal use of crypto currency is focused predominantly on the online market places for the sale and purchase of certain illicit goods and services. The same report indicated that such risk could rise in the future and the current use of digital currency as a method by which terrorists raise or move money out of the UK provides a viable and working method of doing so \cite{cg16}.

Law enforcement serious concerns also include the relative ease in which the perpetrators,
such as the terrorist organizations or organized criminal group operatives,
can increase and enhance their operational security.
They may, for example, use anonymous email services for the purpose of setting-up
the crypto currency wallets and -- for further authentication -- use the anonymous prepaid phones \cite{cg06}.
Although some countries introduced laws which theoretically require the identification and registration of the mobile phone and SIM-card buyers,
it is technically very easy to obtain any number of SIM-cards using the services of the so-called “straw buyers” or “smurfs”,
that is persons who offer their identification documents in exchange for money \cite{cg17},
the alternative being the purchase of virtual, pre-registered SIM-cards.

From the point of view of the law enforcement, the additional enhancement of operational security for the criminal offenders
and terrorists wishing to exploit blockchain technologies in the form of crypto currency is the availability of bitcoin
(or other virtual currency) at Automated Teller Machines (ATMs) and the freestanding electronic payment consoles enabling purchase of and payments in bitcoin.
For example, such un-monitored (lacking any kind of surveillance) consoles are available in Romania – and they may be used with the application of the simple operational security techniques mentioned above (single-use, anonymous e-mail accounts, single-use “burner” mobile phone and prepaid SIM-card) \cite{cg16}.

The examples of the threats that raise the concerns of the law enforcement and counter-terrorism agencies in regards to crypto currencies include the online postings by the supporters of the so-called ISIS (or ISIL: the Islamic State or Iraq and the Levant) that include YouTube videos, discussion forum links and links to research and anonymity provided by bitcoins, where direct references are made to using digital currencies to transfer funds into countries where conventional or traditional methods of financial transactions are difficult because of lack of network capacity or government surveillance and regulation \cite{cg07}. The analysts suggest that Bitcoins were used in a number of successful attacks, such as the bombing attacks conducted by the ISIS-inspired lone wolf on the shopping mall in Jakarta, Indonesia in 2015, and in the November 2015 coordinated terror attacks in Paris. The same authors indicate, that the Internet sites associated with terrorist organizations have started to collect donations in bitcoins \cite{cg07}. Other researchers state explicitly that, given the lack of adequate controls over transactions with bitcoin, the risk of the anonymous transactions between entities financing terrorism is extremely high \cite{cg05}.
Some scholars consider crypto currency to be the next step in the evolution of the more traditional ways of financing terrorism internationally. They refer to the ancient system called Hawala, used in the Middle East and Asia to transfer funds across borders in a safe and convenient manner: funds are deposited with the hawala broker who arranges for the funds to be available from another hawala broker in a different country (both hawala brokers then settle accounts through the normal process of trade). Hawala can be used for terrorism financing and money laundering because funds do not actually cross borders, removing the international money trail \cite{cg07}. The process may be considered slow and inefficient nowadays, but with the development of new technologies, such as crypto currency, it can be significantly accelerated, providing that the contemporary “brokers” (or intermediaries) are trusted, thus minimizing the risk of the authorities learning about the illegal activity.
Some security experts indicate, that – apart from the terrorism financing - crypto currencies are an ideal vehicle for money laundering, bribery and financing of illegal activities. The rationale behind it is to disguise the origins of illicit proceeds throughout a series of transactions prior
to integrating the crime proceeds into the legitimate financial system. In regards to use of crypto currency, the process (divided into three stages of: placement, layering and integration) looks as follows \cite{cg18}:
In the placement stage the crime proceeds are introduced into the financial system by acquiring crypto currencies.
After the crime proceeds have entered the financial system,
the perpetrator engages in a series of transactions to distance the funds from their source. It is the layering stage, where the funds may be channeled through the purchase of crypto currencies or by transferring money electronically through a series of crypto currency accounts. It can be further disguised as payment for goods and services or through the use of intermediaries who purchase crypto currencies under the reporting threshold in countries where such currencies are regulated (to avoid triggering identification or reporting requirements). In the final stage of integration, the “cleaned” money is introduced to the legal market, where it appears to be legally earned \cite{cg18}.
Obviously, such stages of money transfer and money laundering are present also in the more “traditional” forms of introducing the proceeds of crime to the regular legal market, but the aforementioned methods of operational security and the accessibility plus the sheer speed of transactions
facilitated by the use of crypto currencies 
significantly improves the pace and security of such dealings, providing that the persons involved maintain the operational security algorithms at all stages of their behavior.

According to the most recent report by RAND Corporation \cite{cg54},
there is little evidence of the adoption of crypto currencies by terrorist organizations,
and little motivation to do so. This is changing and 
lone-actor extremists and loosely associated groups are now likely to use crypto currencies,
simply because these systems are nowadays very widely used.

\newpage 
\section{
{Crypto Currency Related Crimes: Overview Classification and Stats}}
\label{sec:10Klaus:CrimesClassifyProport}

Our research indicates that criminality involving crypto currency
allows for profits that are much higher than with conventional felonies.
In addition, stealing 
virtual money appears to be 
less risky than robbing a bank or selling drugs.
In this section we will attempt to classify crypto currency crimes in several categories and sub-categories.
We are not the first to propose this type of classification,
see \cite{cg37} which document is no longer publicly available,
and has a broader focus
on policing the whole so called dark web activity.
Our approach is more basic and centered around payment technology: we ``follow the money''.
Our goal is also to achieve some degree or proportionality:
we spend more time on crimes which were prominent in the recent years
and had a large economic or public policy impact, cf. \cite{cg49}.
Our initial short classification has 9 major categories,
which are not entirely disjoint, and which are as follows:
\vskip5pt

\begin{enumerate}
\item
{\bf Crypto Hacks}.
Crypto currency itself is hacked.
This doesn't happen very often but the impact can be enormous.
Major issues can be as follows.
\begin{enumerate}
\item
The peer network malfunctions or attackers
achieve important network advantage (e.g. censoring or undoing transactions).
\item
The integrity and authenticity of digital signatures is broken.
\item
The integrity of blockchain itself is compromised due to a hash function weakness.
\item
Smart contracts or cross-chain swaps or oracle/reference servers fail to work correctly
e.g. in 
DAO heist with 50 M\$ losses 
\cite{DAO}.
\item
The system is secure but implementation is insecure
(e.g. with side channel attacks
the private key of multiple users can be compromised
and all their coins stolen).
\item
There is failure at operation, for example with bad randoms or weak passwords \cite{cg13,cg14}.
\end{enumerate}
\item
{\bf Thefts}. 
A crypto currency exchange or a business entity is hacked or dishonest,
and money is stolen. 
\begin{enumerate}
\item
Indiscriminate loss: the whole exchange is a victim of a theft.
Such events are typically widely publicized, e.g. 
\cite{Tether30M2017} and pages 22-32 in \cite{cg24},
one notable exception was Crex24 hack \cite{cg24}.
\item
Fraudulent withdrawal: a hacker manages to impersonate a user and withdraw his funds.
\item
Fraud at payment deposit stage: 
Server that accepts crypto currency (e.g. online market place) is hacked,
money is stolen.
\item
Client using crypto currency is exposed in interaction or hacked, his money is stolen.
\end{enumerate}
\item
{\bf Investment Fraud@Exchanges.}
\begin{enumerate}
\item
Pure ponzi schemes. For example with OneCoin, PlusToken and Wotoken and similar,
several billions of dollars in deposits were collected from naive investors.
cf. 
\cite{cg24} and \cite{cg50}.
\item
A crypto currency operator deliberately creates barriers or limitations:
disabling withdrawals or deposits, banning certain categories of users,
creating fake technical incidents, etc.
\item
A crypto exchange will list or delist coins on false pretexts in order to manipulate the market.
\item
Promoting alternative coins which have little or no value. Production of fake news about crypto coins.
For example the Wall Street Journal have once written that GAW miners was the world's fastest growing bitcoin mining operation.
Later the founder was charged with defrauding hundreds of individuals around the world for 9 M\$, cf. \cite{GAWscam}.
\item
Dishonest use of information 
provided by users or traders,
e.g. making financial bets against customers.
\end{enumerate}
\item
{\bf Exit Scams.} They can occur {\bf not only} at exchanges.
\begin{enumerate}
\item
Users at a crypto exchange cannot withdraw funds due to arbitrary decisions or on false technical pretexts.
\item
Adopters of a specific crypto currency or ICO investors
see the sellers of this token disappear.
\item
Failure to deliver coins:
customer pays for crypto currency but he doesn't receive any.
\item
With miners: see 5.c.
\end{enumerate}
\item
{\bf Software and Hardware Scams} provide
malicious or fake tools or services: 
\begin{enumerate}
\item
Insecure wallets which expose users to thefts.
\item
Criminals recompile and modify some free open-source software.
\item
Failure to deliver miner equipment paid for by customers.
\item
Miners with backdoors
\cite{MineBackDoorShut}.
\end{enumerate}
\item
{\bf Web Scams}: a fraudulent web site imitating a legit company
which is attracting investor activity e.g. fake exchanges.
\item
{\bf Malware} can steal crypto currency at one of.
\begin{enumerate}
\item
At server/exchange side.
\item
At client wallet side.
\item
Or mining crypto currency at user expense.
\end{enumerate}
\item
{\bf Ransomware/Extortion}.
Crypto currency can be used for ransom payment with blackmail or kidnapping etc.
\item
{\bf Law Enforcement Failures}.
\begin{enumerate}
\item
Police cannot seize crypto currency because the owner does not collaborate,
for example he does not reveal his password.
\item
Crimes are specifically crafted to operate across jurisdiction boundaries
or with unreliable offshore business entities.
\item
Concealing monetary flows from the authorities, e.g.
money laundering with crypto currency.
\end{enumerate}
\end{enumerate}


\subsection{
{Scams, Ransoms and Other Blockchain Fraud}}
\label{sec:10:1:FraudRans}

According to a recent crime report by CipherTrace from February 2021, 
cf. \cite{cg49}, ever since 2019, fraud and misappropriation are the dominant form of crypto currency crime.
Then come the thefts, and finally we have ransomware which comes 3rd.
According to \cite{cg50} scams
account for 54 $\%$, while ransomware accounts for 7 $\%$ of criminal activity.
We cover ransomware in more detail in Section \ref{sec:10:A:Ransom} below.
In 2019 scams have received an enormous 10 billion dollars from naive investors,
cf. \cite{cg24,cg50}.
Another author have identified 132 scams defrauding investors for a total of 16 billions of dollars since 2012, see \cite{Scam16G}.
Interestingly after 2019, in 2020 the scam activity has apparently substantially declined cf. Fig. 1 in \cite{cg49}. 
In 2020-21 we have seen many scams related to Covid.
Scams remain a very major problem at all times.
They concern an extremely large population of more than 200 million amateur crypto investors \cite{CryptoComAdoptReport0721} and beyond.

\subsection{Ransoms Are Back}
\label{sec:10:A:Ransom}

A very major area of concern for the law enforcement agencies and security services is the potential behind the use of crypto currencies in ransomware attacks.
In the famous older case of the CryptoLocker malware of 2013,
the trojan in question was using the 256-bit AES cryptography to encrypt user files so that they became irrecoverable.
Nearly 250,000 individuals and businesses around the world suffered because of the CryptoLocker attack,
which earned an estimated 30 million USD for its developer \cite{cg08} in a period of just 100 days between mid-2013 and May 2014 \cite{cg09}.
Following \cite{cg50} the total amount of ransoms paid in 2019 was about 100 M\$ in one year in 2019,
and this amount has more than tripled in 2020 to reach an estimated 350 M\$.
The ransomware attacks pose the extremely serious threat where they victimize not only the individuals and small business, where the potential ransom is relatively small, but when they target banks, telecoms, health service and critical national infrastructure.
Since CryptoLocker, ransomware offenders tried to attack among others:
the telephone provider TalkTalk in November 2015, many banks etc.
Bitcoin remains the primary method of paying the ransom \cite{cg16}.
Privacy-enhanced coins such as Dash account for less than 2 $\%$, cf. \cite{ransom20}.

A major ransomware attack was certainly the coordinated, worldwide WannaCry assault of May 2017.
It infected 230,000 computers in 150 countries
and the total amount of losses was estimated 
at 4 billion USD \cite{ransom20}.
Among the targeted institutions were the Deutsche Bahn, FedEx,
National Health Service (NHS) in the UK, 
governmental and Police institutions in several countries,
airlines, universities, automobile manufacturers, etc.
The attack was stopped within a few days of its discovery due to emergency patches released by Microsoft,
and the discovery of the kill switch that prevented infected computers from spreading WannaCry further.
A preliminary evaluation by security experts
stated that the attack originated from North Korea \cite{cg55}.
Then in 2018 and 2019 ransomware has declined, to explode with more force in 2020, cf. page 31 in \cite{cg50}.
In 2020 the most prominent victims of ransomware were among other Barnes$\&$Noble,
LG or University Hospital of New Jersey \cite{cg50}.
A significant observation in recent and major attacks, is that criminals publish some of the data online in order
to increase the pressure on the victims \cite{cg50}.
For example, in February 2021, the source code of the blockbuster video game maker CD Project was stolen.
The company declined to pay the ransom.
Then reportedly the code was then auctioned on dark markets for 7M \$, see \cite{cg51}.

This resurgence of ransoms in 2020 is not at all surprising and seems to be a logical consequence of two trends.
The first is that losses are reported in dollars and clearly the bitcoin price has increased up to 8  times in 2020 making some older ransom strains  programmed to request certain standardized amounts in bitcoins substantially more profitable.
A second observation is that,
with the global Covid pandemic,
businesses are more desperate to maintain business continuity at all costs, and thus more inclined to pay ransoms,
for example Travelex reportedly paid 2.3 million in ransom, see \cite{cg24}.
In 2020 we have seen malicious Covid-related Android applications which spied on users, encrypted files and asked for ransoms, cf. page 23 in \cite{cg24}.
Peer to peer financial systems are a substantial challenge for the law enforcement and security agencies and have a set of characteristics making them very attractive to the criminal and terrorist enterprises \cite{cg06}: a certain level of anonymity (if the necessary operational security precautions are implemented), global reach (being geo-political border agnostic \cite{cg20} and allowing to carry on the transactions through third countries), the speed
(facilitating the quick transactions thus limiting the chance of them being intercepted or blocked), low cost to use (dependent of the fees of the intermediaries in case of the crypto currency transfer, tumbling and exchange), relative ease of use (being increasingly user friendly – as it is true with most of the emerging technologies, TOR being the prime example), difficulty to track by the authorities, and the lack of legal regulations and control over the decentralized crypto currency ledgers servers and distributed or collaborative electronic systems.

\subsection{On Kindnapping Front}
\label{sec:10:C:Kindap}

There are several high-profile kidnapping cases where ransoms in millions of dollars were paid.
For example, in 2017, a criminal group kidnapped an employee of a United Kingdom-registered crypto currency exchange in Kiev, Ukraine.
The kidnappers released the victim on 29 December, reportedly following a ransom payment equivalent to USD 1 million in Bitcoin.
Several similar events were reported and we refer \cite{KindapThreat} for a comprehensive review of this topic.
It appears that crypto currency ransoms are not the norm or not yet for kidnappings in general.
Instead, kidnappers are targeting wealthy individuals in general.
It appears that in several cases of kidnapping where crypto currency payments were requested,
the kidnappers were able to competently exploit open or public information
to identify individuals, whose crypto currency wealth have been publicly known.

\section{
{New Criminal Battlefield Revisited}}
\label{sec:11:NewxCryptoCrimBattlefiled}

In 2011, when the notion of the “Internet as a new criminal battlefield” was first introduced \cite{cg21},
the blockchain concept and the crypto currencies were virtually unknown outside of the specialists and crypto enthusiasts’ circles.
However, for these and many other technological developments, the use of any technology for the purposes that are (directly or potentially) illegal in nature, is somewhat normal, as there is no reason why criminals would not adopt a new tool 
especially if it is readily available, easy to use and inexpensive.
Additionally, the younger generation of criminals we are dealing with are more likely to embrace the Internet and technology
and will consider it to be their natural environment.
Therefore, we can assume that the frequency of on-line based and cyber-enabled crime increases in parallel to the generational changes.
It needs to be stressed that the Internet and new technology are used at all stages of criminal behavior.
From the planning and preparation, through the completion of an act, destroying or altering the evidence,
up to the stage of preparing an alibi or manipulating witnesses \cite{cg21}.
We could compare these technologies to a multi-purpose tool, which was not originally designed to cause harm,
but which was later skillfully and creatively adapted,
by those who found other and previously unforeseen applications for it.
Now we should not ignore the fact that certain very specific types of crimes, and this is {\bf primarily ransomware},
are specifically enabled by this new technology.
In the “old world”, it was harder for criminals to receive payments without being caught or monitored.
Blockchain technology also offers the possibility to defer the moment at which the criminal takes the money for potentially unlimited time,
and this is a completely new scenario that has never existed before.
Therefore we must also recognize that beyond being just a tool, blockchain can genuinely be a disruptive technology
and modify the landscape of criminal activity in which criminals engage.

A useful comparison here is to the credit card payment history: there are several countries including the UK, France and Malaysia
where the introduction of Chip and Pin technology \cite{cg12} was motivated partly by the quick increase of crime and fraud rates.
In the same way, the very possibility of using credit cards to purchase goods over the Internet has played a major role in crime,
and created a pre-bitcoin situation where stealing data has become a potential moneymaker.
Bitcoin has just amplified the necessity to innovate, in the areas of payment and the digital economy,
and is the next logical step after the credit cards.
First we had credit cards since the beginning of the 20th century, cf. \cite{cg12},
then in 1990s the credit cards have become trustless,
in the sense that they possess private keys which authenticate transactions
and they are the only entity able to authenticate payments
so that our money is safe against fraud even when 
we do not trust the bank.
It is easy to see that even the bank cannot forge
or alter transactions made with Chip and Pin credit cards.
Finally we have various forms of disintermediation.
First of all payment companies such as VISA, Mastercard or PayPal
have deprived the banks from substantial amount of income from payments,
and the income goes to the multinational firms, rather than traditional high street banks.

Then we have a new wave of disintermediation, which seems ``anarchic'' like bitcoin.
Strangely, money lands at the end in some large crypto exchanges
which are based in the United States primarily cf. \cite{cg49}.
Businesses of that kind have enjoyed phenomenal growth rates.
Being a fan or a user of bitcoin currency, in spite of undeniable extremely large
impact and popularity \cite{CryptoComAdoptReport0721},
is possibly
somewhat {\bf a smoke screen} which hides a broader business reality.
A large number of new unregulated para-bank companies are competing for ``geek'' customers
which are fans of bitcoin and other innovations,
and any method to capture their attention and their deposits is good.
Disintermediation and business competition across borders 
is the main pattern.
We simply observe banks losing business and customers to certain offshore businesses.
Crypto exchanges can now also compete wtih banks on getting funding from investors.
In 2021, a major crypto currency exchange, coinbase,
has successfully become a regular traded company on Nasdaq stock exchange
with a market cap of 70 billion approximately at the time of writing,
which is comparable to some of the largest banks in the world.

Bitcoin is just a part of an inevitable trend in which payment technology and financial markets must change
to serve the needs of the global digital economy better, and to respond and adapt to what the new technology can enable and offer.
We have a continuous wave of business disruption, where jobs and business are moving to different jurisdictions.
This is hardly avoidable. The crime is evolving likewise, taking new and unprecedented forms

An interesting point, is that in parallel, blockchain creates a new liberal ``geek'' offshore jurisdiction
where maths and code are law. Interestingly it is also a self-policing jurisdiction.
It is meant to regulate itself or to work in very different ways than the incumbent banking sector.
We need to recognise that our old financial system was very costly to run, for example the police forces of the whole
planet collaborate on a very large scale in order to police fake currency crime. This problem disappears with bitcoins,
as un-forge-ability is a standard feature in modern cryptography.
It is not true that the security is systematically degraded or that the lack of regulation is systematically a problem.
On one side, bitcoin facilitates certain specific forms of crime such as ransoms, cf. Section \ref{sec:6:B:SHA256Centr}
or certain types of transgressions are allowed or tolerated in the crypto economy.
Then on the other side, security is reinforced at many places. Accordingly, many crimes do not happen anymore.
Many attacks or thefts occur only in very specific circumstances and typically one just cannot take bitcoins which are not theirs,
because of strong cryptography, see Section \ref{sec:6:A:TrustCry} and
Section \ref{sec:6:B:SHA256Centr}.

\subsection{Law Enforcement Shortcomings}
\label{sec:11:A:ShortCom}

A key problem is that the law enforcement is radically unprepared to deal with the new types of payment business and supra-national business disruption, due to
-- among others -- the administrative and legal constraints,
lack of funding (translating in majority of the countries to the inability to compete financially with the private sector
when it comes to hiring the properly trained and skilled IT experts), lack or expertise and training, ineffective or non-existent exchange of information, inability or difficulty in cross-border cooperation, the general lack of knowledge and experience in regards to cyber- and cyber-enabled threats \cite{cg03}. According to our recent interviews with the Intelligence and security practitioners, the problem of the abuse of modern technologies by criminals and terrorist requires the multi-disciplinary approach involving the co-operation of the law enforcement, criminal justice, computer scientists and the industry \cite{cg03}. No matter how cliché it might sound, the most significant improvement in the current situation can probably be achieved just by raising awareness and education of all parties involved. This specifically relates to the law enforcement community. Qualitative studies performed during the EC FP7 PRIME Project \cite{cg03}, \cite{cg17} research show that the level of knowledge about modern technologies among the Police officers worldwide is extremely limited. One of the objectives of this paper is to make the computer science community and IT industry aware of this fact. Law enforcement professionals have a very vague understanding of the emerging technologies and their potential for abuse by the criminal and terrorist enterprises. Moreover, we have unrealistic expectations about the potential and skills of the Police among the general public and computer science industry representatives. The so-called “CSI Effect” originally related to the impossibility to achieve (the alleged) “standards”
in forensic sciences is currently starting to impact the popular expectations towards “cyber investigations”,
resulting in naïve and improbable anticipations about the skills and capabilities of the Police forces.
It is thus necessary to re-consider and re-work the existing strategies and methodologies of Police education. No matter how idealistic it might sound, in order to do it, it is necessary to involve the computer science industry in such a framework, as the rapid changes and developments in technology render the traditional law enforcement training based on the well-established and long-term curriculum largely useless.
A highly plausible approach \cite{cg03} is to engage in public-private partnerships, with law enforcement practitioners,
lawyers, computer science experts and cyber-forensics specialists.
We recommend using the approach developed by the field of Crime Science, namely the crime-scripting,
understood as a way of deconstructing an offence into its basic component actions, and also criminal horizon scanning \cite{cg22}, \cite{cg23}.
Recent developments in the area are the new EU-funded project Titanium developing blockchain analytic tools
named GraphSense and DarkNet Monitor.
Interpol has a working group 
and they have published at intepol.int web site
a draft taxonomy under development of crimes involving 
darknets and crypto currencies, see \cite{InterpolDark}.

\subsection{Difficulties and Perspectives}
\label{sec:11:B:DiffPers}

Finally, it is important to re-consider the future of the legal framework in which the well-established, emerging and forthcoming technologies will operate. As mentioned earlier \cite{cg20}, cyber-enabled criminal or terrorist acts are geopolitically border agnostic, as their perpetrators can operate simultaneously in several jurisdictions, and the flexibility and potentially high anonymity available to them hamper the potential for prevention, detection, and prosecution. From the legal standpoint, the international harmonization and unification of the “cyber law” is a slow process. In theory, it is more likely in the case of operational and procedural domain (including problem prevention, identification, and recognition; risk detection and assessment; evidence investigation, gathering, and presentation; and crisis management regulations), although – as our findings imply \cite{cg03} -- numerous substantial discrepancies still exist.

\subsection{Restoring Audit  with Focused AML}
\label{sec:11:C:FocusAML}

Money laundering cannot be just policed by banning some cryptographic inventions such as ZeroCash, but rather by “focused AML”, term used by a US law firm in \cite{cg35}. In a focused AML approach we disclose additional private transaction information to our bank or exchange, but not to a larger crowd.
For example, many systems such as Stealth Adress in Monero or ZeroCash have so called “view keys” for this purpose.
The fact is that these keys enable to establish high standards of proof.
They are cryptographically unforgeable,
and potentially they have more value than current AML reporting standards at crypto exchanges,
where documents submitted could for example be fake, or authentic which were carefully crafted to show an inaccurate picture.

\section{
{The New Economy is Here and it Deserves a Legal Protection  }} 
\label{sec:12:NewEcoNeedsLegalProt}

In the discussion of the problem of the Internet as a new battlefield,
we see that law enforcement and the criminal offenders compete for domination \cite{cg20}.
At the same time technology solutions and financial services compete for the sake of business,
with well-known moral hazards, namely compromising public safety and security
in order to acquire a larger market share.
It was noted that although it might seem natural that such issues should be a domain of politicians and legislators,
the design of universal strategies and intervention measures is actually in the hands of scientists, industry, and end-users (security services). A central role is in fact going to be played by new emerging disciplines and new emerging professions, which are all in sense also new chapters of modern Computer Science. In particular we have the topics of Big Data, Applied Cryptography, Artificial Intelligence and Deep Learning.
We see that anonymity and innovation is disruptive for law enforcement sector as much as they are for the banking and payment industry.
We can anticipate that the law enforcement sector needs to change their culture radically in the sense that they need to become knowledge-based,
employ radically more technicians, research engineers, data scientists, and specialists with PhDs in related fields.
Following \cite{cg44} criminal activities involving crypto currencies require to develop novel analytic methodology and tools.
Apart from the cultural shift, it is very difficult to expect the international legal community to come up with a comprehensive vision or strategy, when dealing with challenges related to emerging technologies and business and software ecosystems which lead to radical changes in criminal landscape. This is a process, and it requires collaboration on the international scale \cite{cg37}.
As of today, we do not yet have an all-inclusive and up-to-date, universally agreed set of solutions \cite{cg37}. One of the reasons for this is that different nations have very different levels at which they have embraced new technology and cyber-security. Another problem is that different nations have different views on free speech, civil liberties such as usage of private communication or privacy enhancing technology, and personal or consumer rights and freedoms. Some countries will be tempted to address the emerging cybercrime problems from the point of view of nations, which stay on the top of new technology implementation. Other nations will try to avoid complication and maybe say: “forbid or regulate the payment technology we do not understand or master well enough”. Instead of having the philosophical or typical problem of maintaining the delicate balance between two fundamental expectations: the right to privacy which is recognized as a human right, and the right to safety which is more about how industrial nations protect their citizens and businesses. We observe that different nations approach the problem differently due to their culture and history.

\subsection{On Risks of Banning or Blacklisting}
\label{sec:12:B:RiskBann}

Money laundering cannot be just policed by banning some powerful cryptographic inventions such as ZeroCash.
We need to embrace new technology rather than see each new technology development as an anomaly, which we cannot handle.
Banning is in fact hurting business, innovation and the 99.9 $\%$ of users.
In \cite{cg38} authors study payments from the point of view of network topology and connectivity and define fitness as the ability of the node to attract new connections. Then they show that the accounts that have high fitness are of two sorts.
They are either short-lived and indulge in malicious or criminal activities, or they are long-lived and they represent large organizations.
Here is the dilemma: policing crime without negatively impacting businesses.

\newpage

\section{
{Perspectives of Regulating and Policing Financial Innovation Provider Businesses}}
\label{sec:13:ReguCryptBusi}

We need a pro-innovation and pro-business agenda: whether the new technologies such as blockchain are just a “hype” or not, nations are trying to attract the innovators to create jobs in their countries. In some sense new technology is always an uncertain bet from the point of view of investing in it,
and assessing how successful or how influential they might become.
If so, it is perfectly logical for the countries to embrace disruptive innovations,
maintaining their original paradigm of independence and freedom, and hoping that some unicorn companies
are going to become rich and pay taxes,
in a friendly jurisdiction which figures out the right mix of freedom to innovate and regulation.
Such philosophy was attempted in New York with Bit License and in France with AMF licencing.
Then we deal with the associated criminal activity as it comes, with the idea of proportionality.
In contrast, over-regulating and planning to be ten steps ahead of criminals,
is an expensive strategy for law enforcement agencies and security services.

\subsection{Can Safeguards which Prevail In Financial Regulation Survive in the Immaterial World? }
\label{sec:14:SafegImmat}

We need to look at the spirit of laws and regulations which have underpinned the payment and see that many ideas are obsolete with new technology, and also with how modern capitalism works at large.
The whole idea is that a company can issue a “security” and this process must be very heavily policed by lawmakers and law enforcement
agencies such as SEC in order to protect the investors is potentially obsolete.
Since 2000 numerous financial scandals and investment bubbles have shown that investors
do and will fall into traps
and buy the 
trendy share or coin,
which is not always even trying to do a good job as an investment.
Even properly regulated securities on the stock markets
have failed to protect millions of investors in practice.

The crucial question here is that issuing a security is heavily regulated,
and is expected to be a reliable link between what investors expect to own or invest in,
and some material business operations or assets.
We want investors to know what they are buying and investors are entitled to be treated fairly
and get the same share of corporate profits in proportion, as for example the founders of the company.
However the whole idea of business assets being material and subject to policing by accountants,
have become totally obsolete in the recent years.
Now the 5 top tech companies on S+P500 stock index are worth more than 20 \% of the whole index,
and the combined value of US tech companies in general is more than the whole stock market in Europe.
The key observation is that a great part of what we see on the stock market
are immaterial assets and intangible property,
or assets which are non-exclusive.
An Internet platform or system can benefit potentially anyone,
and it is not possible to monopolize the virtual wealth stemming from the network effects,
the economies of scale and collaboration easily. In this light, it is wrong to believe that
bitcoin benefited only some scammers or speculators.
On Internet forums we observe that there is a broad population of people,
who are not unhappy with the profitability of their crypto
currency investments.
In some sense investors were right to invest in the Internet bubble in 2000
and in Tesla stocks or Bitcoin and Ethereum today.
They understood that tremendous amounts of wealth, power and innovation are at stake,
and are going to be created somewhere here, for example with new electric cars.

In fact it remains difficult to capture, own or monopolize all this wealth.
Stock markets and government regulation of how securities are issued have
frequently and again, failed to protect against investors investing in wrong assets.
Some crypto currency enthusiasts contend that crypto currency companies
will soon be allowed to issue securities and they will be fully compliant with US Laws and regulations.
This is no happening yet and so far such companies have been facing severe regulatory backlash.
In July 2021 many companies (e.g. Binance, UniSwap) have suddenly halted trading crypto tokens which
get too close into imitating stock market shares.
Arguably however, one day, maybe in 10 or 20 yeas, securities could be eventually issued on blockchains.
This is because blockchains can provide important and extensive new functionality,
which does not exist in current financial markets.

In our view, 
both technology and law enforcement should work in their own way,
on improving technical standards and safeguards.
Ultra conservative blacklisting, banning or fear mongering attitudes,
are rarely compatible with disruptive new technology which creates new jobs, new ways of doing things,
or new ways to secure property or engage in business contracts, and this is how our economy evolves of improves.
Crypto market regulations should protect the small to average investors,
primarily not against taking active part, and effectively co-funding the development of new disruptive technology,
but against rogue traders, rogue developers, faulty cryptography standards,
market manipulators, and predatory market mass surveillance
leading to targeted fraud and crime
activities.


\newpage 

\section{
{Conclusion}}
\label{sec:conclusion}

Financial crime or tax avoidance are serious problems,
which however should not conceal the fact that it is hardly possible to
forbid users from using alternative unregulated financial services.
Large software ecosystems such as Bitcoin or Ethereum blockchains are nowadays considered
as good investment opportunities, and have year after year gained an immense popularity.
Bitcoin did not go to zero, instead it has crossed the 60,000 \$ mark.
Many crypto currencies have high prices due to artificially created scarcity
and poor investment opportunities elsewhere.
However there is no scarcity of developers bringing a plethora
of innovative DeFi and crypto protocols to market.
The blockchain ledgers grow every day and occupy hundreds of gigabytes.
There is no scarcity of transactions either.
As of July 2021, blockchain has 200 million of users \cite{CryptoComAdoptReport0721}.
Every single day 1 million bitcoin wallets are in use.

\subsection{Actionable Recommendations}

We need to uphold the free market economy 
while protecting investors from predatory forces and thieves through
better privacy, better cryptography and better cybersecurity.
If we look at the figures we see that
the biggest threats to individual customers are not thefts or ransoms,
but primarily scams which simply imitate
the investment opportunities in the new economy \cite{cg50,Scam16G}.
Then, comes ransomware, a major threat on the rise, where victims are primarily businesses.
Crypto currency coins offer some unique features such decentralized ownership
and programmable decentralized trading which brings to us, a completely new world of
business functionalities. 

A controversial feature is privacy for all parties involved in financial transactions.
Arguably future blockchain and cybersecurity standards should operate on the principle of proportionality
and tackle the actual cyber crime rather than ban and forbid in advance.
Proportionality would suggest to enable more privacy in the financial sector:
it benefits the 99.9$\%$ of legitimate activity.
This even though it also helps the 0.1 $\%$ of illicit activity which remains small \cite{cg49}.
According to a recent paper by a specialist US lawyer firm \cite{cg35}
the societal or economic benefits from privacy coins are substantial
and “substantially outweigh their risks”.
According to this law firm, the situation is broadly under control,
and existing AML regulations already
“properly and sufficiently cover" those risks w.r.t. money laundering. 

\section*{
{Acknowledgements}}

Although the preparation of this paper was not directly funded
by a designated research grant,
our research and expertise have been supported indirectly from different grants
and by our universities.
Nicolas Courtois has also worked as a consultant
for several crypto currency companies in UK, France and the US.
His background and inspiration in his security engineering and payment security research and teaching
were eight years spent working as a crypto R$\&$D engineer for the French smart card industry
(Bull CP8, now part of Gemalto), and seven years of interactions and engagement with the crypto currency community.
He was also funded by the EU with project RIBS on Resilient Infrastructure and Building Security.
Kacper Gradon acknowledges the support of UCL Jill Dando Institute of Security and Crime Science.
Some of the ideas expressed here arouse during the preparation of the deliverables
of the EU-funded 
PRIME Project and throughout the duration of the U.S.
State Department Fulbright Scholarship at Institute of
Behavioral Science University of Colorado at Boulder.

\end{document}